\makeatletter
\@ifundefined{@parse@version@dash}{%
\def\@parse@version#1{\@parse@version@0#1}
\def\@parse@version@#1/#2/#3#4#5\@nil{%
\@parse@version@dash#1-#2-#3#4\@nil}
\def\@parse@version@dash#1-#2-#3#4#5\@nil{%
  \if\relax#2\relax\else#1\fi#2#3#4 }
}{}
\makeatother

\documentclass[reprint,amsmath,amssymb,aps,showkeys]{revtex4-2}

\usepackage{float}
\usepackage{caption}
\usepackage{subcaption}
\usepackage{graphicx}
\usepackage{dcolumn}
\usepackage{bm}
\usepackage{booktabs}

\begin{document}
	
\title{Self-consistent on-site and inter-site Hubbard parameters within DFT+U+V for UO$_2$ using density-functional perturbation theory}
\author{Mahmoud Payami }
\email{mpayami@aeoi.org.ir}
\author{Samira Sheykhi}
\author{Mohammad-Reza Basaadat}
\affiliation{School of Physics \& Accelerators, Nuclear Science and Technology Research Institute, AEOI,\\ 
	P.~O.~Box~14395-836, Tehran, Iran}

\begin{abstract}
To apply the Hubbard-corrected density-functional theory for predicting some known materials' properties, the Hubbard parameters are usually so tuned that the calculations give results in agreement with some experimental data and then one uses the tuned model to predict unknown properties. However, in designing new unknown novel materials there is no data to fit the parameters and therefore self-consistent determination of these parameters is crucial. In this work, using the new method formulated by others, which is based on density-functional perturbation theory, we have calculated self-consistently the Hubbard parameters for UO$_2$ crystal within different popular exchange-correlation approximations. The calculated ground-state lattice constants and electronic band-gaps are compared with experiment and shown that PBE-sol lead to results in best agreement with experiment.   
\end{abstract}


\maketitle

\section{Introduction}\label{sec1}

In applying the density-functional theory (DFT) \cite{hohenberg1964,kohn1965self} to calculate a material's properties, practically one uses approximations for the exchange-correlation (XC) energy functional. Among the employed approximations, the most popular ones, i.e., local-density approximation (LDA) \cite{kohn1965self,perdewzunger81} and the generalized gradient approximation (GGA) \cite{gga-pbe1996} suffer from self-interaction errors, which limits their applicabilities to weakly-correlated electron materials bonded with {\it s} and {\it p} orbitals. In systems containing atoms with localized {\it d} and {\it f} orbitals, this error is higher and lead to over-delocalization of their corresponding wave-functions which in turn lead to incorrect prediction of metallic behavior for Mott insulators. Different methods have been proposed to cure this problem. One of the methods, which is computationally very expensive, is using variants of hybrid orbital-dependent functionals among them HSE is computationally low-cost \cite{linlin2016,SHEYKHI201893}. Another method, which is most popular and computationally very low-cost compared to the latter method, is using Hubbard model to correct the correlation energy of localized orbitals in DFT energy-functional. The simplest such corrected method, called the DFT+U, adds only on-site corrections to the DFT energy functional \cite{coco-degironc2005}: 

\begin{equation}\label{eq1}
E_{\rm DFT+U}=E_{\rm DFT}[n({\bf r})] + E_{\rm Hub}[{n_m^{I\sigma}}]-E_{\rm dc}[n^{I\sigma}],
\end{equation}
in which $n({\bf r})$ is electron density, $n_m^{I\sigma}$ are occupation numbers of orbitals of atom at lattice site ${\bf R}_I$, and $n^{I\sigma}=\sum_m n_m^{I\sigma}$. The last term in right hand side of Eq.~(\ref{eq1}) is added to avoid double counting of interactions contained in the first and second terms.
The simplified rotationally invariant form \cite{dudarev1998} of the correction is given by \cite{coco-degironc2005}: 

\begin{equation}\label{eq2}
\begin{split}
E_{\rm U}[{n_{m m^\prime}^{I\sigma}}]\equiv E_{\rm Hub}-E_{\rm dc}\;\;\;\;\;\;\;\;\;\;\;\;\;\;\;\;\;\;\;\;\; \\
         \;\;\;\;\;\;\;\;=\sum_{I,\sigma}\frac{U^I}{2} {\rm Tr[{\bf n}^{\it I\sigma}({\bf 1} - {\bf n}^{\it I\sigma}})],
\end{split}
\end{equation}
in which ${\bf n}^{I\sigma}$ is the atomic occupation matrix. This correction, which is on-site correction, significantly corrects the incorrect prediction of metallic behavior of Mott insulators.

In the extended method called DFT+U+V, which in addition to on-site corrections, includes the inter-site corrections, the rotationally invariant correction to DFT energy functional becomes \cite{campojr2010}:

\begin{equation}\label{eq3}
\begin{split}
E_{\rm U+V}\equiv E_{\rm Hub}-E_{\rm dc}=\sum_{I,\sigma}\frac{U^I}{2} {\rm Tr[{\bf n}^{\it II\sigma}({\bf 1} - {\bf n}^{\it II\sigma}})] \\
-\sum_{I,J,\sigma}^*\frac{V^{IJ}}{2} {\rm Tr[{\bf n}^{\it IJ\sigma}{\bf n}^{\it JI\sigma}}],\;\;\;\;\;
\end{split}
\end{equation}
in which the asterisk $*$ over the sum in second term indicates that the $J$ sums are over the atoms lying on the spherical concentric shells with atom $I$ at the center. That is, the 1st shell contains the nearest neighbor atoms, the 2nd shell contains the next to nearest neighbor atoms, etc.
The coefficients $U^I$ and $V^{IJ}$ are called Hubbard parameters. 

When we are dealing with a known material, these parameters may be empirically so tuned that the calculations give results in agreement with some experimental data and then employing the tuned model one is able to predict unknown properties of that material. In a new recent work \cite{payami-lattice-eg}
it was shown that taking into account only the on-site parameters for $5f$ orbital of uranium and $2p$ orbital of oxygen atoms in UO$_2$, and tuning their values for the LDA approximation of XC and choosing projection operator onto non-orthogonalized Hubbard orbitals, it is possible to reproduce the experimental values of lattice constant as well as the electron band gap. It was also shown that this can happen for different pairs of parameter values indicating a degrees of freedom to select one that also lead to some third correct experimental data. However, in designing new unknown novel materials there is no data to fit the parameters. Additionally, it is advantageous to have a parameter-free theory and therefore self-consistent determination of these parameters is crucial. 

In the earlier attempts to self-consistent determination of Hubbard parameters, linear-response constrained-DFT (LR-cDFT) were used within super-cell method \cite{coco-degironc2005,campojr2010}. The method was somewhat cumbersome and computationally expensive. The new method which was introduced in 2018, uses density-functional perturbation theory within the unit-cell, which is relatively fast and also feasible with low-memory computational facilities \cite{tim2018,tim2021,tim2022}. 
The self-consistent parameters are determined using HP code \cite{tim2022} included in the Quantum-ESPRESSO code package \cite{qe-2009,qe-2020}. In this work, using the HP code, we have done a series of self-consistent calculation of Hubbard parameters in DFT+U and DFT+U+V schemes for UO$_2$ crystal in the contexts of LDA \cite{perdewzunger81} and gradient-corrected variants: PW91 \cite{pw91}, PBE \cite{gga-pbe1996}, PBE-sol \cite{gga-pbesol2008}, and rev-PBE \cite{revpbe2008}. The calculations include results for both "atomic" and "ortho-atomic" types of projections. 

This paper is organized as follows: Section~\ref{sec2} provides the computational details; in Section~\ref{sec3} the calculated results are presented and discussed; and Section~\ref{sec4} summarizes the conclusions of this research.

\section{Computational details}\label{sec2}
\begin{figure}
	\centering
		\includegraphics[width=0.3\textwidth]{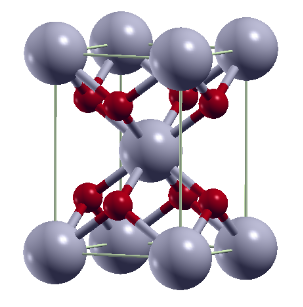}
	\caption{UO$_2$ crystal structure as simple tetragonal with six atoms basis. Large gray and small red balls represent uranium and oxygen atoms, respectively.}
	\label{fig1}
\end{figure}

For the description of crystal structure we have used a simple tetragonal lattice with a six-atoms basis, shown in Fig.~\ref{fig1}. In addition, to setup anti-ferromagnetic (AFM) structure for U atoms, we used the simple model in which the planes of U atoms alternate their spins when moving in $z$ direction, i.e., a 1-dimensional AFM.  

\begin{figure}
	\centering
		\includegraphics[width=0.3\textwidth]{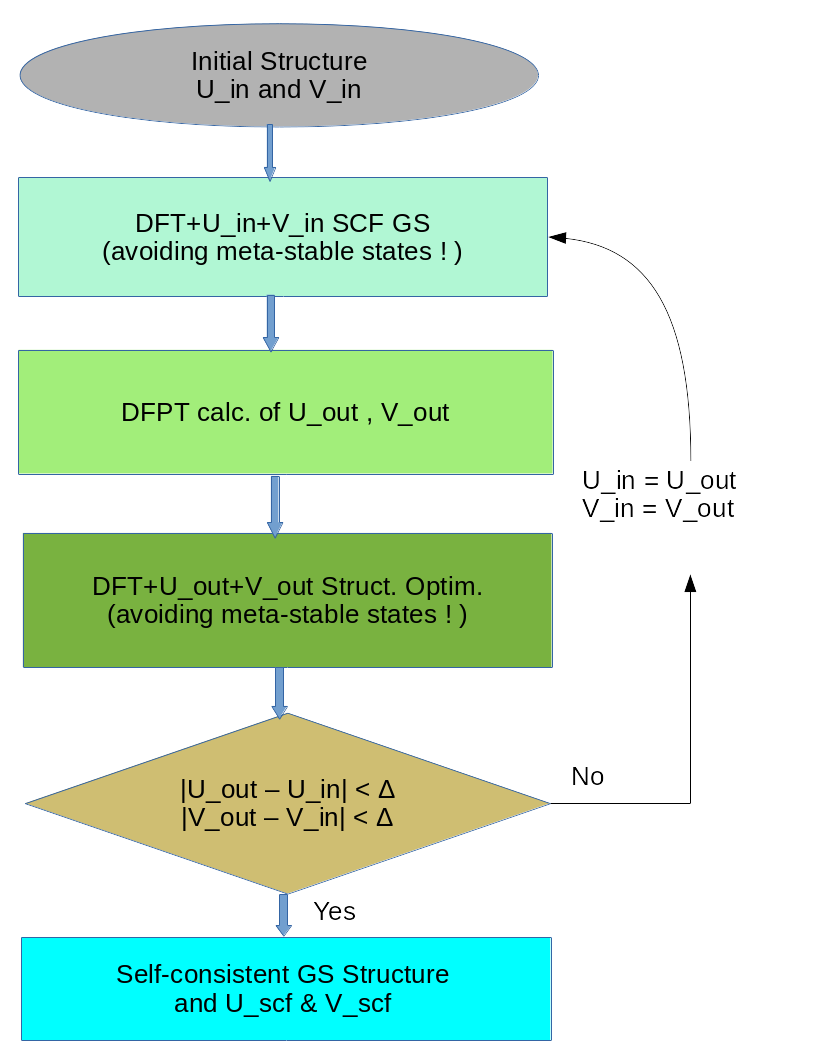}
	\caption{Flowchart of SCF determination of Hubbard parameters. In the first-step SCF and last-step structure-optimization, the meta-stable states were avoided \cite{payami-spinbroken2021,payami-smcomc2023}}
	\label{fig2}
\end{figure}

Determined the Hubbard parameters using the HP code, all DFT, DFT+U, and DFT+U+V calculations were based on the solution of the KS equations using the Quantum-ESPRESSO code package \cite{qe-2009,qe-2020}. For U and O atoms we have used scalar-relativistic ultra-soft pseudo-potentials (USPP) generated by the {\it atomic} code and generation inputs from the {\it pslibrary} \cite{DALCORSO2014337}, at https://github.com/dalcorso/pslibrary. The valence configurations U($6s^2,\, 6p^6,\, 7s^2,\, 7p^0,\, 6d^1,\, 5f^3 $) and O($2s^2,\, 2p^4 $) were used in the USPP generation.
Different approximations for the XC interactions were used: LDA \cite{perdewzunger81} and gradient-corrected variants: PW91 \cite{pw91}, PBE \cite{gga-pbe1996}, PBE-sol \cite{gga-pbesol2008}, and rev-PBE \cite{revpbe2008}. 
The calculations include two cases when the Hubbard atomic orbitals were taken as orthogonalized or not.
Kinetic energy cutoffs for the plane-wave expansions
were chosen as 90 and 720~Ry for the wave-functions and densities, respectively. The smearing method of Marzari-Vanderbilt \cite{mv-smear1999} for the occupations with a width of 0.01~Ry were used. 
For the Brillouin-zone integrations in geometry optimizations, a $8\times 8\times 6$ grid were used;  All geometries were fully optimized for total residual pressures on unit cells to within 0.5 kbar, and residual forces on atoms to within 10$^{-3}$~mRy/a.u.
To self-consistent determination of the Hubbard parameters we have employed the HP code \cite{tim2022} following the flowchart shown in Fig.~\ref{fig2}. 

To start the self-consistent procedure for Hubbard parameters according to Fig.~\ref{fig2}, we give initial values for $U_{in}$ and $V_{in}$ to open the electronic bang-gap; for the initial structure we choose simple tetragonal structure with appropriate lattice constants consistent with cubic structure of side 5.47$\AA$. To avoid meta-stable states, we choose appropriate occupations of Hubbard orbitals $5f$ of uranium atoms \cite{payami-spinbroken2021,payami-smcomc2023}. Our experience shows that starting from metalic ground state lead to divergence of some components of response function. 
For Hubbard-corrected DFT calculations, we consider on-site corrections for only $5f$ orbitals of uranium atoms, and the inter-site corrections for $5f$ orbitals of U atoms and $2p$ orbitals of first nearest neighbor O atoms. In the second step, we start the DFPT calculation and obtain new values for parameters named as $U_{out}$ and $V_{out}$. In the third step, using the parameters $U_{out}$ and $V_{out}$ obtained in the second step, we optimize the geometry of the system taking care of meta-stable states. In each cycle we monitor the differences between input and output parameters to see if the self-consistency is reached within $\Delta$ value. For this system the self-consistency is reached within 6 to 8 cycles in the flowchart.  

\section{Results and discussions}\label{sec3}
The calculations were done at three levels of approximations: i)-simple DFT with no Hubbard corrections, ii)-DFT with on-site Hubbard corrections (DFT+U), and iii)-DFT with on-site and inter-site corrections (DFT+U+V). For the simple DFT calculations we obtain incorrect metalic behavior, while for other two cases we get insulating properties with different geometric and electronic properties. The results are presented in Table~\ref{tab1}.

\begin{table*}\centering
\caption{Caption}
\begin{tabular}{@{}lcccccccccccc@{}}\toprule
& \multicolumn{2}{c}{DFT} & \phantom{abc}& \multicolumn{3}{c}{$\;\;\;\;\;\;\;\;\;\;\;\;$DFT+U} &
\phantom{abc} & \multicolumn{4}{c}{DFT+U+V}\\
\cmidrule{2-3} \cmidrule{6-8} \cmidrule{10-13}
XC& $a\;(c)\;(\AA)$ & $E_g\;(eV)$ & U-proj && $U\;(eV)$ & $a\;(c)\;(\AA)$ & $E_g\;(eV)$ && $U\;(eV)$ & $V\;(eV)$ & $a\;(c)\;(\AA)$ & $E_g\;(eV)$\\ \midrule
LDA &5.286\;(5.300) &metal & atomic &&2.14&5.414\;(5.443)&1.18 && 2.22 & 0.37&5.406\;(5.437)&1.18\\
    &  &  & ortho && 3.11 & 5.414\;(5.431) & 2.04 && 3.09 & 0.21&5.410\;(5.430) & 2.02\\ \midrule
PW91&5.472\;(5.487) &metal & atomic &&1.98&5.517\;(5.530)&1.61 && 2.01 & 0.32&5.509\;(5.524)&1.59\\
    &  &  & ortho && 2.96 & 5.516\;(5.517) & 2.50 && 2.88 & 0.18&5.513\;(5.515) & 2.46\\ \midrule
PBE&5.472\;(5.450) &metal & atomic &&1.99&5.519\;(5.533)&1.57 && 2.02 & 0.32&5.512\;(5.527)&1.56\\
    &  &  & ortho && 2.96 & 5.518\;(5.520) & 2.46 && 2.89 & 0.17&5.516\;(5.518) & 2.43\\ \midrule
PBE-sol&5.337\;(5.353) &metal & atomic &&2.06&5.454\;(5.476)&1.35 && 2.11 & 0.31&5.447\;(5.471)&1.36\\
    &  &  & ortho && {\bf 3.03} & {\bf 5.454}\;({\bf 5.463}) & {\bf 2.27} && {\bf 2.91} & {\bf 0.18}&{\bf 5.451}\;({\bf 5.460}) & {\bf 2.19}\\ \midrule
rev-PBE&5.457\;(5.463) &metal & atomic &&1.99&5.550\;(5.554)&1.63 && 2.01 & 0.32&5.543\;(5.548)&1.62\\
    &  &  & ortho && 2.96 & 5.545\;(5.546) & 2.49 && 2.84 & 0.15&5.542\;(5.543) & 2.42\\ 
\bottomrule
\end{tabular}\label{tab1}
\end{table*}

As is seen from Table~\ref{tab1}, the DFT calculations for PW91, PBE, and rev-PBE give good lattice constants comparable with experiment but incorrect metalic properties. On the other hand, all Hubbard corrected results show correct insulating properties. Among all Hubbard-corrected calculations, the ones obtained using PBE-sol approximation for the XC are in good agreement with experiments. For both DFT+U and DFT+U+V, the non-orthogonalized Hubbard atomic orbitals lead to band gaps far from the experimental value. On the other hand, using the orthogonalized orbitals lead to band gaps of 2.27 eV for DFT+U and 2.19 eV for DFT+U+V in good agreement with experiment \cite{schoenes1978optical}. Taking into account the inter-site correction does not significantly modify the geometry, but lowers the band gap by 0.08 eV. In Table~\ref{tab1} the results for ortho-atomic PBE-sol are bolded.

\section{Conclusions}\label{sec4}
In the study of strongly-correlated materials with some known experimental properties, it is not uncommon that the Hubbard perameters be chosen in such a way that lead to results consistent with known experimental properties and then continue the calculations with those tuned parameters to predict other unknown properties. However, the latter method is not applicable in designing new novel materials because of the lack of experimental investigations and one has to determine those Hubbard parameters self-consistently. 
In this work, using DFPT we have determined the self-consistent Hubbard parameters for different XC approximations and have shown that PBE-sol with ortho-atomic projection gives the best results.  

\section*{Acknowledgement}
This work is part of research program in School of Physics and Accelerators, NSTRI, AEOI.  

\section*{Data availability }
The raw or processed data required to reproduce these results can be shared with anybody interested upon 
sending an email to M. Payami.

\end{document}